\documentclass[preprint,aps,newabstract]{revtex4}

\usepackage{graphicx}
\usepackage{axodraw}
\usepackage{latexsym}

\newcommand{\lsim}{\mathrel{\mathop{\kern 0pt \rlap
  {\raise.2ex\hbox{$<$}}}
  \lower.9ex\hbox{\kern-.190em $\sim$}}}
\newcommand{\gsim}{\mathrel{\mathop{\kern 0pt \rlap
  {\raise.2ex\hbox{$>$}}}
  \lower.9ex\hbox{\kern-.190em $\sim$}}}
\newcommand{\beq}     {\begin{equation}}
\newcommand{\eeq}     {\end{equation}}
\newcommand{\bea}     {\begin{eqnarray}}
\newcommand{\eea}     {\end{eqnarray}}

\newcommand{\al}      {\alpha}

\newcommand{\n}      {{(n)}}

\newcommand{\es}      {\epsilon}

\newcommand{\Gam}      {\Gamma}

\newcommand{\M}       {{\mathcal M}}

\newcommand{\no}      {\nonumber}
\newcommand{\hats}      {\hat{s}}
\newcommand{\longto}    {-\!\!\!-\!\!\!\longrightarrow}

\newcommand{\nc}{\newcommand}
\nc{\postscript}[2]
{\setlength{\epsfxsize}{#2\hsize}\centerline{\epsfbox{#1}}}
\nc{\bg}{B. Grzadkowski}
\nc{\non}{\nonumber}
\nc{\hc}{\hbox {h.c.}} \nc{\re}{\hbox {Re}} 
\nc{\mev}{\hbox {MeV}} \nc{\gev}{\;\hbox {GeV}} \nc{\tev}{\;\hbox {TeV}}
\def\lsim{\mathrel{\raise.3ex\hbox{$<$\kern-.75em\lower1ex\hbox{$\sim$}}}}
\def\gsim{\mathrel{\raise.3ex\hbox{$>$\kern-.75em\lower1ex\hbox{$\sim$}}}}

\nc{\Lsp}{\;\;\;\;\;\;\;\;\;\;}  \nc{\LLLsp}{\lspace \lspace}
\nc{\lsp}{\;\;\;\;\;\;}
\nc{\spac}{\;\;\;}
\nc{\noi}{\noindent}
\nc{\baa}{\begin{array}}      \nc{\eaa}{\end{array}}
\nc{\bit}{\begin{itemize}}    \nc{\eit}{\end{itemize}}
\nc{\ben}{\begin{enumerate}}  \nc{\een}{\end{enumerate}}
\nc{\bce}{\begin{center}}     \nc{\ece}{\end{center}}

\def\Hhat{\widehat H}

\def\ho{h_0}
\def\mho{m_{\ho}}
\def\phio{\phi_0}
\def\mphio{m_{\phio}}

\def\gam{\gamma}

\def\lwh{\widehat\Lambda_W}

\def\lphi{\Lambda_\phi}

\def\hbar{\overline h}
\def\grav{h_{\mu\nu}^{(n)}}
\def\lam{\lambda}
\def\Lam{\Lambda}
\def\mpl{M_{\rm Pl}}
\def\ifmath#1{\relax\ifmmode #1\else $#1$\fi}

\def\eps{\epsilon}

\begin{document}

\renewcommand{\thepage}{-- \arabic{page} --}
\def\mib#1{\mbox{\boldmath $#1$}}
\def\bra#1{\langle #1 |}      \def\ket#1{|#1\rangle}
\def\vev#1{\langle #1\rangle} \def\dps{\displaystyle}
\nc{\tb}{\stackrel{{\scriptscriptstyle (-)}}{t}}
\nc{\bb}{\stackrel{{\scriptscriptstyle (-)}}{b}}
\nc{\fb}{\stackrel{{\scriptscriptstyle (-)}}{f}}
\nc{\pp}{\gamma \gamma}
\nc{\pptt}{\pp \to \ttbar}
\nc{\barh}{\overline{h}}
   \def\thebibliography#1{\centerline{REFERENCES}
     \list{[\arabic{enumi}]}{\settowidth\labelwidth{[#1]}\leftmargin
     \labelwidth\advance\leftmargin\labelsep\usecounter{enumi}}
     \def\newblock{\hskip .11em plus .33em minus -.07em}\sloppy
     \clubpenalty4000\widowpenalty4000\sfcode`\.=1000\relax}\let
     \endthebibliography=\endlist
   \def\sec#1{\addtocounter{section}{1}\section*{\hspace*{-0.72cm}
     \normalsize\bf\arabic{section}.$\;$#1}\vspace*{-0.3cm}}
\preprint{ \hbox{\bf hep-ph/0311295}}
\preprint{ \hbox{NSC-NCTS-031101}} \preprint{ \hbox{KIAS-P02085}}

\vskip 1cm

\title{Probing the Radion-Higgs mixing at hadronic colliders}

\author{
Kingman Cheung\footnote{cheung@phys.nthu.edu.tw}}
\affiliation{Department of Physics and NCTS, National
Tsing Hua University, Hsinchu, Taiwan}
\author{
C.~S. Kim\footnote{cskim@yonsei.ac.kr}}
\affiliation{Department of Physics and
IPAP, Yonsei University, Seoul 120-749, Korea}
\author{
Jeonghyeon Song\footnote{jhsong@konkuk.ac.kr}}
\affiliation{Department of Physics, Konkuk University,
                   Seoul 143-701, Korea}

\begin{abstract}

\noindent In the Randall-Sundrum model,
the radion-Higgs mixing is weakly suppressed
by the effective electroweak scale.
One of its novel features
would be a sizable three-point vertex of $\grav$-$h$-$\phi$.
We explored the potential of the Fermilab Tevatron and the CERN LHC
in probing the radion-Higgs mixing via the associated production of
the radion with the Higgs boson.
The observation of the rare decay of the KK gravitons into  $h \phi$
is then the direct and exclusive signal of the radion-Higgs mixing.
We also studied all the partial decay widths of the KK gravitons in
the presence of the radion-Higgs mixing, and found that
if the mixing parameter is of order one,
the decay rate into a radion and a Higgs boson
becomes as large as that into a Higgs boson pair,
with the branching ratio of order $10^{-3}$.

\end{abstract}

\maketitle

\section{Introduction}

The standard model (SM) has been
extraordinarily successful
in explaining all experimental data on the electroweak
interactions
of the gauge bosons and fermions up to now.  However,
the master piece of the SM,
the Higgs boson, still awaits
experimental discovery\,\cite{Higgs}.
Theoretical consideration of the triviality and the unitarity puts
an upper bound of $ (8\pi \sqrt{2}/3 G_F)^{1/2}\sim 1$
TeV\,\cite{trivial,Duncan:1986vj} on the Higgs boson mass. 
On the other hand,
the direct search has put a lower mass limit of 114.4 GeV on the SM
Higgs boson at the 95\% C.L.\,\cite{higgs}, while
the indirect evidences from
the electroweak precision data imply a light
Higgs boson of order ${\mathcal O}(100)$ GeV\,\cite{lepew}.
In order to establish the Higgs mechanism for the electroweak
symmetry breaking, one
also requires to study in detail the Higgs boson interactions
with the gauge bosons and fermions.
Therefore,
one of the primary goals of future collider experiments is directed
toward the study of the Higgs boson.

The Higgs boson is also a clue
to various models of new physics beyond the SM,
as its mass receives radiative corrections
very sensitive to the UV physics.
This is the so-called gauge hierarchy problem.
Recently, a lot of theoretical and phenomenological interests
have been drawn to a scenario proposed
by Randall and Sundrum (RS) \cite{RS},
where an additional spatial dimension of a $S^1/Z_2$ orbifold is introduced
with two 3-branes at the fixed
points.  A geometrical suppression factor, called the warp factor,
emerges and
naturally explains the huge hierarchy between the electroweak and Planck scale
with moderate values of the model parameters.
A stabilization mechanism was introduced \cite{GW} to maintain the
brane separation, and to avoid unconventional
cosmological phenomenologies \cite{Csaki-cosmology}.
Such a mechanism introduces a radion much lighter
than the Kaluza-Klein (KK) states of any bulk fields.
In the literature,
various phenomenological aspects of the radion
have been studied such as
its decay modes\,\cite{Ko,Wells},
its effects
on the electroweak precision observations\,\cite{Csaki-EW},
and its phenomenological signatures
at present and future colliders\,\cite{collider}.

In the viewpoint of the Higgs phenomenology,
the presence of another scalar (the radion)
may modify the characteristics of the Higgs boson itself
even in the minimal RS scenario where all the SM fields
are confined on the TeV brane.
It is due to the radion-Higgs mixing originated
from the gravity-scalar mixing term,
$\xi R(g_{\rm vis}) \widehat{H}^\dagger \widehat{H}$,
where $R(g_{\rm vis})$ being the Ricci scalar of the induced metric
$g_{\rm vis}^{\mu\nu}$.
Here $\widehat{H}$ is the Higgs field in the five-dimensional context.
It has been shown that the radion-Higgs mixing can induce significant
deviations to the properties of the SM Higgs
boson\,\cite{Datta-HR-LHC,Han-unitarity,Hewett:2002nk,Gunion}.

A complementary way to probe the radion-Higgs mixing
is the direct search for the new couplings
exclusively allowed with a non-zero mixing parameter $\xi$.
One good example is the tri-linear vertex among
the KK graviton, the Higgs boson, and the radion.
In Ref.~\cite{ours},
we have shown that, especially in the limit of large VEV of the radion,
probing the $h_{\mu\nu}^{(n)}$-$h$-$\phi$
through the $h \phi$ production at $e^+ e^-$ colliders
can provide very useful information on the radion-Higgs mixing,
irrespective of the mass spectrum of the Higgs and radion.
This high energy collision process
is complementary to the rare decay modes of the Higgs boson
allowed with non-zero $\xi$, $e.g.$,
$h \to \phi\phi$ which can be sizable in some
parameter space\,\cite{Gunion}.

In this work, we focus on the associated production of
the radion with the Higgs boson at hadronic colliders,
the Fermilab Tevatron and the CERN Large Hadron Collider (LHC).
The higher center-of-mass (c.m.) energy of the hadron colliders
allows the on-shell production of KK gravitons,
the decay of which in turn yields clean signals of the RS model.
The observation of the rare decay of the KK gravitons into  $h \phi$
is then the direct and exclusive signal of the radion-Higgs mixing.
In addition, the characteristic
angular distribution could reveal the exchange of massive spin-2 KK
gravitons.

This paper is organized as follows.
Section \ref{sec:review} summarizes
the RS model and the basic properties of the radion-Higgs mixing.
In Sec.~\ref{sec:G-gam}, we calculate the partial decay
widths of the graviton in
the presence of the radion-Higgs mixing.
The production cross section of
$p \,p\, (\bar{p}) \to h \,\phi$ and the corresponding
kinematic distributions are discussed in Sec.~\ref{sec:production}.
Section \ref{sec:detection}
deals with the feasibility of detecting the $h \phi$ final states
by considering specific decay channels of the Higgs and radion.
We summarize and conclude in
Sec.~\ref{sec:conclusion}.

\section{Review of the Randall-Sundrum model and radion-Higgs mixing}
\label{sec:review}

The RS scenario is based on a five-dimensional spacetime
with non-factorizable geometry\,\cite{RS}.
The single extra dimension is compactified on a
$S^1/Z_2$ orbifold of which two
fixed points accommodate two three-branes,
the Planck brane at $y=0$ and
the TeV brane at $y=1/2$.
Four-dimensional Poincare invariance is shown to be maintained
by the following classical solution
to the Einstein equation:
\beq
ds^2=e^{-2\sigma(y)}\eta_{\mu\nu}dx^\mu dx^\nu-b_0^2dy^2,
\label{metricz}
\eeq
where
$\eta_{\mu\nu}$ is the Minkowski metric,
$\sigma(y)=m_0 b_0|y|$, and $y \in [0,1/2]$.
The five-dimensional Planck mass $M_5$ ($\eps \equiv 1/M_5^3$)
is related to the four-dimensional Planck mass
($M_{\rm Pl} \equiv 1/\sqrt{8\pi G_N}$) by
\beq
{M_{\rm Pl}^2\over 2}={1-\Omega_0^2  \over \eps^2 m_0},
\label{mplkappam0}
\eeq
where $\Omega_0\equiv e^{-m_0 b_0/2}$ is the warp factor.
On the TeV  brane one observes the mass of
a canonically normalized scalar field to be
multiplied by the small warp factor, $i.e.$, $m_{phys}=\Omega_0 m_0$.
As the moderate value of $m_0 b_0/2 \simeq 35$
can generate TeV scale physical mass,
the gauge hierarchy problem is answered.

In the minimal RS model,
all the SM fields are confined on the TeV brane.
Gravitational fluctuations
about the RS metric such as
\beq
\eta_{\mu\nu} \to \eta_{\mu\nu}+\epsilon h_{\mu\nu}(x,y)
,\lsp b_0\to b_0+b(x)\,
\label{metric}
\eeq
yield two kinds of
new phenomenological ingredients on the TeV brane,
the
 KK graviton mode $h_{\mu\nu}^{(n)}(x)$
and the canonically normalized radion field $\phi_0(x)$, defined by
\beq
h_{\mu\nu}(x,y) = \sum_{n=0}^\infty
h_{\mu\nu}^{(n)}(x)\frac{\chi^{(n)}(y)}{\sqrt{b_0}}
\,, \quad
\phio(x) = \sqrt{6}\mpl\Omega_b(x)\,,
\eeq
where $\Omega_b(x)\equiv e^{-m_0[b_0+b(x)]/2}$.
The four-dimensional effective Lagrangian is then
\beq
{\cal L}=-\frac{\phi_0}{\Lambda_\phi} T_\mu^\mu
-\frac{1}{\lwh}  T^{\mu\nu}(x) \sum_{n=1}^\infty
h^{(n)}_{\mu\nu}(x)
\,,
\eeq
where $\Lambda_\phi(=\sqrt{6}\mpl \Omega_0)$
is the vacuum expectation value (VEV)
of the radion field, $T^\mu_\mu$ is the trace of the symmetric
energy-momentum tensor $T^{\mu\nu}$,
and $\lwh = \sqrt{2} \mpl \Omega_0$.
Note that both effective interactions are suppressed by the
electroweak scale, not by the Planck scale.

All the SM symmetries and Poincare invariance on the TeV brane
are still respected by the following gravity-scalar
mixing term~\cite{Wells,Gunion}:
\beq
S_\xi=\xi \int d^4 x \sqrt{g_{\rm vis}}\,R(g_{\rm vis})\Hhat^\dagger \Hhat\,,
\eeq
where $R(g_{\rm vis})$ is the Ricci scalar for the induced metric
on the visible brane,
$g^{\mu\nu}_{\rm vis}=\Omega_b^2(x)(\eta^{\mu\nu}+\eps h^{\mu\nu})$,
$H_0=\Omega_0 \Hhat$, and
$\xi$ denotes the size of the mixing term.
This $\xi$-term mixes the $h_0$ and $\phi_0$ fields
into the mass eigenstates of $h$ and $\phi$ fields, given by\,\cite{Gunion}
\beq
\label{matrix}
\left(
\begin{array}{c}
  h_0 \\
  \phi_0 \\
\end{array}
\right)
=
\left(
\begin{array}{cc}
  1 & 6 \xi \gamma/Z \\
  0 & -1/Z \\
\end{array}
\right)
\left(
\begin{array}{rc}
  \cos\theta & \sin\theta \\
  -\sin\theta & \cos\theta \\
\end{array}
\right)
\left(\begin{array}{c}
  h \\
  \phi \\
\end{array}
\right) =
\left(
\begin{array}{cc}
  d & c \\
  b & a \\
\end{array}
\right)
\left(\begin{array}{c}
  h \\
  \phi \\
\end{array}
\right)
\,,
\eeq
where
\bea
Z^2 &\equiv& 1+6\xi\gam^2(1-6\xi)\equiv \beta-36\xi^2\gam^2\,.
\label{z2} \\
\tan2\theta
&=&
12 \gam \xi Z {\mho^2\over \mphio^2-\mho^2(Z^2-36\xi^2\gam^2)}
\,.
\eea
Note that in the RS scenario
the radion-Higgs mixing is
only suppressed by the electroweak scale of $1/\Lam_\phi$.

The eigenvalues for the square of masses are
\beq
m_\pm^2={1\over 2 Z^2}\left\{\mphio^2+\beta \mho^2\pm\sqrt{
(\mphio^2+\beta \mho^2)^2-4Z^2\mphio^2\mho^2}\right\}
\label{emasses}
\,,
\eeq
where $m_+~(m_-)$ is the larger (smaller) between the Higgs mass $m_h$ and the
radion mass $m_\phi$.
Our convention is that in the limit of $\xi\to 0$,
$m_{h_0}$ is the Higgs mass.
The following constraint on $\xi$ from the positivity of the mass  squared
in
Eq.~(\ref{emasses}) is crucially operating in most of parameter space:
\beq
{m_+^2\over m_-^2}>1+{2\beta\over Z^2}
\left(1-{Z^2\over\beta}\right)
+{2\beta\over Z^2}\left[1-{Z^2\over \beta}\right]^{1/2}\,.
\label{rootconstraint}
\eeq

All phenomenological signatures of the RS model
including the radion-Higgs mixing are specified by
five parameters
\beq
\label{parameter}
\xi,\quad \lphi,\quad \frac{m_0}{\mpl},\quad m_\phi,\quad
m_h
\,,
\eeq
which in turn determine $\lwh={\lphi}/\sqrt{3}$ and
KK graviton masses $m_G^{(n)}=x_n {m_0}{\lwh}/({\mpl} {\sqrt{2}})$
with $x_n$ being the $n$-th root of the first order Bessel function.
Some comments on the parameters in Eq.~(\ref{parameter}) are in order.
First, the dimensionless coefficient of the radion-Higgs mixing,
$\xi$, is generally of order one with the constraint in
Eq.~(\ref{rootconstraint}).
The $\Lam_\phi$ which fixes the masses and effective couplings
of KK gravitons is also constrained, $e.g.$,
by the Tevatron Run I data of Drell-Yan process
and by the electroweak precision data:
$m_G^{(1)}\gsim 600$ GeV yields $\lphi\gsim 4$ TeV\,\cite{RS-onoff}.
For
the reliability of the RS solution,
the ratio $m_0/\mpl$
is usually taken around $0.01 \lsim m_0/\mpl \lsim 0.1$
to avoid too large bulk curvature\,\cite{Hewett-bulk-gauge}.
Therefore, we consider the case of $\lphi=5$ TeV and $m_0/\mpl = 0.1$,
where
the effect of radion on the oblique parameters
is small\,\cite{Csaki-EW}.
The radion mass is expected to be light
as one of the simplest stabilization mechanisms
predicts $m_{\phi_0} \sim \lwh/40$\,\cite{GW}.
In addition, the Higgs boson mass is set to be $120$ GeV through out the paper.

\section{Radion-Higgs mixing and Graviton Partial Decay Widths}
\label{sec:G-gam}

The gravity-scalar mixing,  $\xi\, R\, \Hhat^\dagger \Hhat$,
modifies the couplings among the $h$, $\phi$ and $\grav$.
In particular, a non-zero $\xi$ newly generates
the following tri-linear vertices:
\beq
\label{4vertices}
\grav\,\mbox{-}\,h\,\mbox{-}\,\phi,\quad
\grav\,\mbox{-}\,\phi\,\mbox{-}\,\phi,\quad
h\,\mbox{-}\,\phi\,\mbox{-}\,\phi, \quad
\phi\,\mbox{-}\,\phi\,\mbox{-}\,\phi
\,.
\eeq
Focused on the phenomenologies at hadron colliders,
we are interested in the KK graviton production and
its decay exclusively allowed to the radion-Higgs mixing
through the vertices of $\grav-h-\phi$ and
$\grav-\phi-\phi$, defined by
\beq
\langle \, h \, | \,\grav \, | \,\phi \rangle \equiv
i \hat{g}_{G h\phi} \frac{2 k_{1\,\mu}k_{2\,\nu}
}{\lwh}, \quad
\langle\,\phi \,| \,\grav \,|\, \phi \rangle \equiv
{i}\hat{g}_{G \phi\phi}
\frac{2 k_{1\,\mu}k_{2\,\nu}
}{\lwh}\,.
\eeq
Since the parameter $\gam \equiv v_0/\lphi$
is very small with $\lphi=5$ TeV and
\beq
\hat{g}_{G h\phi}
\stackrel{ \gamma \ll 1} {\longto}
{\mathcal O}(\gamma),
\quad
\hat{g}_{G \phi \phi}
\stackrel{ \gamma \ll 1} {\longto}
{\mathcal O}(\gamma^2),
\eeq
$\hat{g}_{G h\phi}$
is much larger than $\hat{g}_{G \phi\phi}$.
Detail expressions are given in Eq.~(\ref{eq:ghat}).
In summary the channel $\grav \to h \phi$ is the most effective in
probing the radion-Higgs mixing.

In the model of RS, graviton KK states are clean resonances.  Therefore,
the production cross section of $p \bar p \to \grav \to h \phi$
depends
critically on the {\em width} of the KK graviton.  At the
graviton pole, the cross section can be expressed as
\begin{equation}
\label{eq:sig-gam}
\hat \sigma (q\bar q,\,gg \to \grav \to h \phi)
\sim \frac{ 8 \pi \Gamma(\grav \to q \bar q,\, gg )\,
   \Gamma( \grav \to h \phi)
       }{ \Gamma^2_{\rm total} m^2_{\grav} } \;,
\end{equation}
where $\Gamma( \grav \to X)$ represents the partial decay width of
$\grav$ into the channel $X$, and $\Gamma_{\rm total}$ is the total decay
width of the graviton.

We calculate all the partial decay widths of the graviton $\grav$ as
a function of its mass in the presence of the mixing $\xi$.
This is a
new result in that the rare decay modes of $\grav \to h \phi, \phi\phi$
and the mass of decay product are taken into account.  The
partial decay widths are given as follows:
\begin{eqnarray}
\Gamma(\grav \to W^+ W^-) &=&
\frac{13}{240\pi}\frac{m_G^3}{\hat \Lambda_W^2}
 \, \left( 1 + \frac{56}{13}\,\mu_W^2
             + \frac{48}{13}\,\mu_W^4 \right )\,
    \sqrt{ 1 - 4 \mu_W^2 }, \\
\Gamma(\grav \to ZZ) &=&
\frac{13}{480\pi} \frac{m_G^3}{\hat \Lambda_W^2}
 \, \left( 1 + \frac{56}{13}\,\mu_Z^2
             + \frac{48}{13}\,\mu_Z^2 \right )\,
    \sqrt{ 1 - 4 \mu_Z^2 }, \\
\Gamma(\grav \to \gamma\gamma) &=& \frac{1}{40\pi}
         \frac{m_G^3}{\hat \Lambda_W^2}, \\
\Gamma(\grav \to gg ) &=& \frac{1}{5\pi}
         \frac{m_G^3}{\hat \Lambda_W^2}, \\
\Gamma(\grav \to f \bar f) &=& \frac{N_f}{80\pi} \,
         \frac{m_G^3}{\hat \Lambda_W^2}
 \, \left( 1 + \frac{8}{3}\,\mu_f^2 \right )\,
    ( 1 - 4 \mu_f^2 )^{3/2},\\
\Gamma(\grav \to h h  ) &=&
\frac{\hat{g}^2_{Ghh}}{480\pi}
         \frac{m_G^3}{\hat \Lambda_W^2} \,
 (  1 - 4 \mu_h^2 )^{5/2}, \\
\Gamma(\grav \to h \phi  ) &=&
\frac{\hat{g}^2_{G h \phi}}{ 240\pi}\,
         \frac{m_G^3}{\hat \Lambda_W^2} \, \beta
  \left[ 1 - \left( \mu_h+\mu_\phi \right )^2 \right ]^2 \,
  \left[ 1 - \left( \mu_h-\mu_\phi  \right )^2 \right ]^2,  \\
\Gamma(\grav \to \phi \phi  ) &=&  \frac{\hat{g}^2_{G \phi \phi}}{ 480\pi}\,
         \frac{m_G^3}{\hat \Lambda_W^2} \,
  ( 1 - 4\mu_\phi^2)^{5/2},
\end{eqnarray}
where $f=q,\ell,\nu_\ell$, $N_f=3 \,(1)$ for $f=q \,(\ell,\nu_\ell)$,
$\mu_x = m_x/m_{G}$,
$\lambda(a,b,c)=a^2+b^2+c^2-2 a b -2 a c-2 b c$ and
$\beta=\lambda^{1/2}(1, \mu_\phi^2, \mu_h^2)$.
Note that the partial decay width depends on the KK graviton mass,
not on the KK mode number.
The total width of the graviton can be obtained by adding all the
partial widths.

\begin{figure}[t!]
\includegraphics{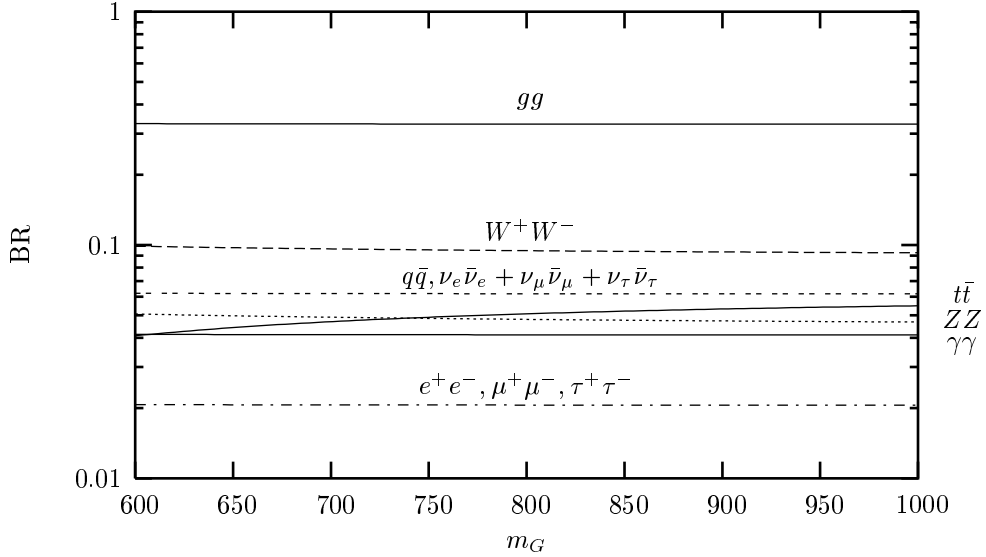}
\caption{\label{gbr} The branching ratios of the KK graviton as a
function of $m_G$ with $\Lam_\phi=5$ TeV. Here $q$ denotes a quark
except for the top quark.}
\end{figure}

In Fig.~\ref{gbr}, we present the branching ratios (BR) of the KK graviton as a
function of its mass. The Tevatron I constraint of
$m_G^{(1)} \gsim 600$ GeV \cite{teva1}  suppresses most of $\mu_i(\equiv
m_i/m_{G} )$--dependence,
implying each BR keeps almost constant.
Only the BR into a top quark pair has moderate dependence on $m_{G}$.
It is clearly shown that the dominant decay mode is into a gluon pair.
The next dominant mode is into
$W^+ W^-$, followed by the modes into a light quark pair and into neutrinos.
The decay into a Higgs pair is suppressed.

\begin{figure}[t!]
\centering
\includegraphics[height=6cm]{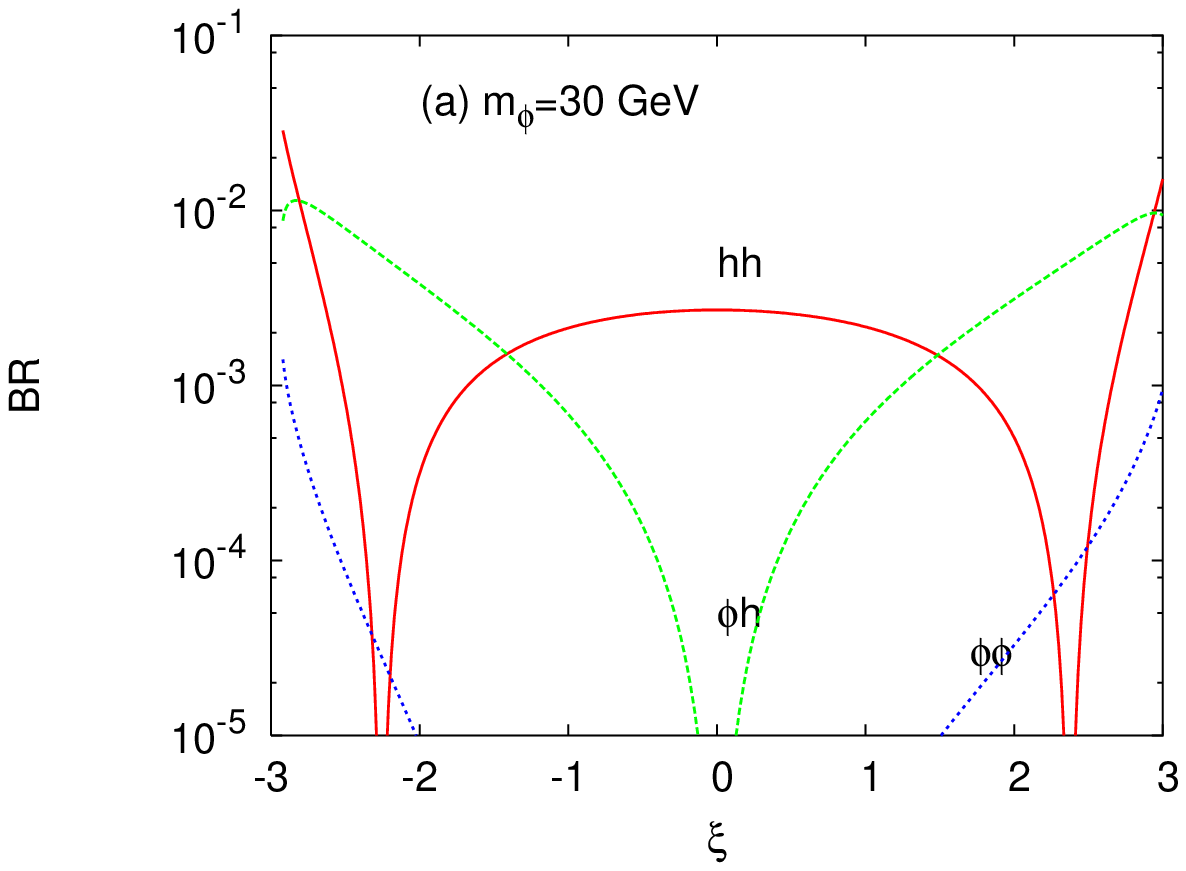}
\includegraphics[height=6cm]{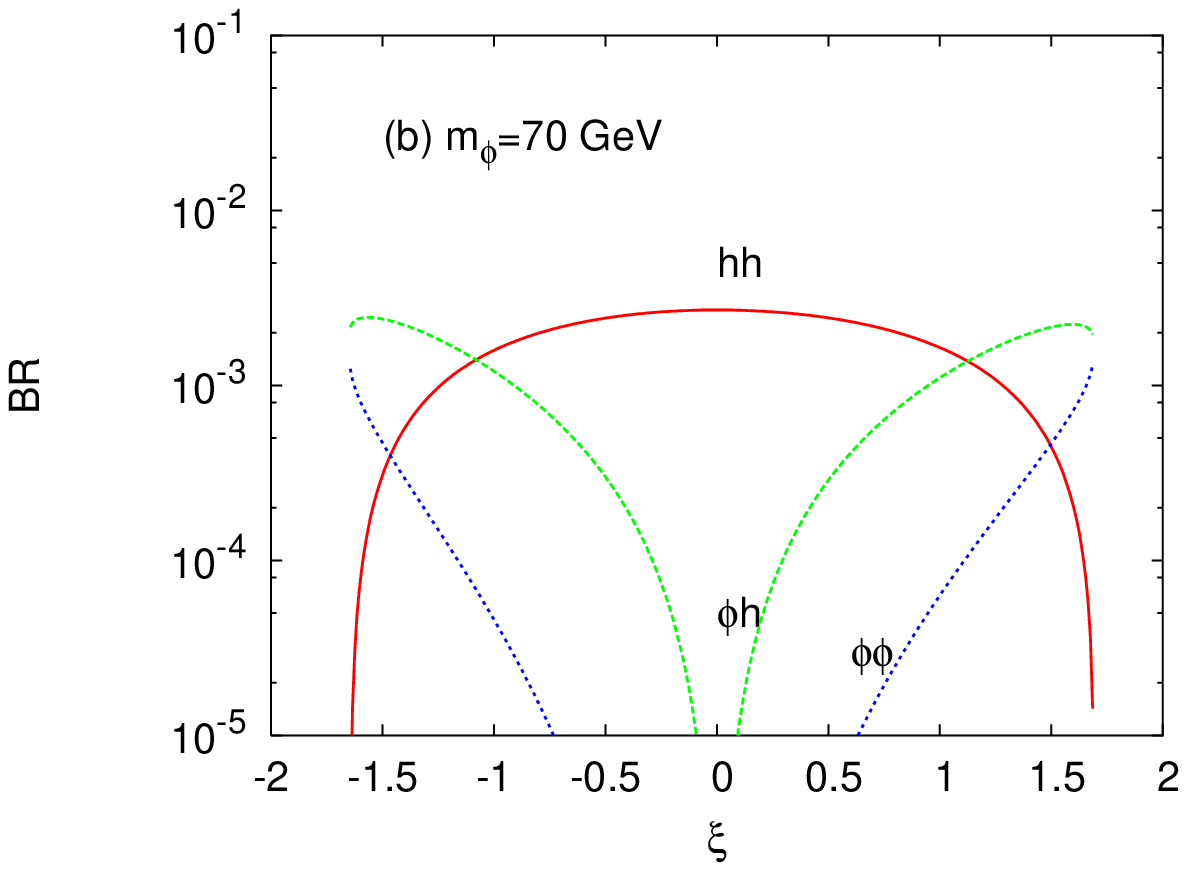}
\includegraphics[height=6cm]{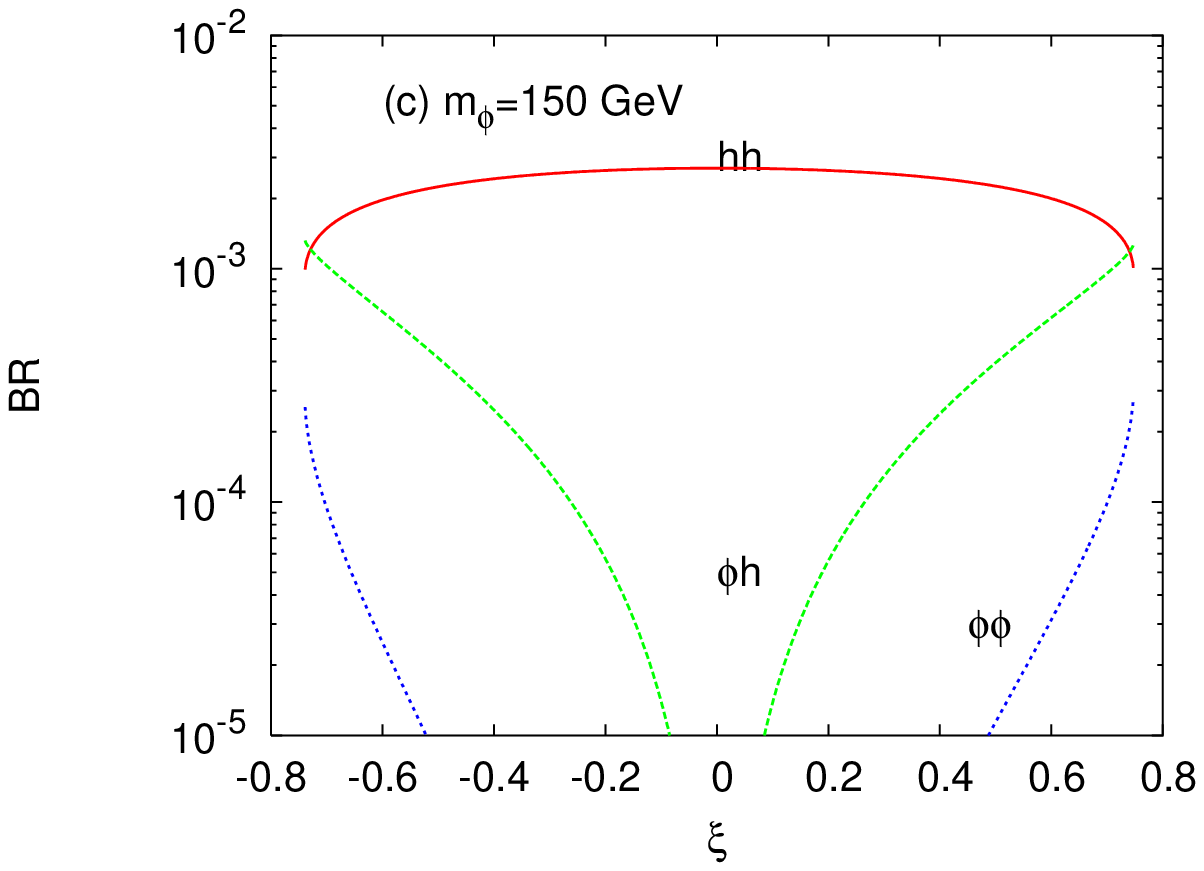}
\caption{\label{xigbr} The branching ratios of the first KK graviton
into two scalars as a function of $\xi$ for (a) $m_\phi=30$ GeV, (b) 70 GeV,
and (c) 150 GeV.  We set $m_h=120$ GeV, $\Lam_\phi=5$ TeV and $m_0/M_{\rm Pl}=0.1$. }
\end{figure}

In Fig.~\ref{xigbr},
we  show the small BR's for
$h_{\mu\nu}^{(1)} \to h h, h\phi, \phi\phi$ as a function of $\xi$ for
$m_\phi=30,~70,~150$ GeV, respectively. As discussed before,
we set $\Lam_\phi=5$ TeV
and $m_0/M_{\rm Pl}=0.1$. Note that if
the radion-Higgs mixing is absent ($i.e.$, $\xi=0$),
$h_{\mu\nu}^{(1)} \to h \phi, \phi \phi$ modes  disappear.
If  $\xi \sim {\mathcal O}(1)$, the BR($\grav \to h\phi$)
for a light radion becomes compatible with BR($\grav \to h h$),
which is of order ${\mathcal O}(10^{-3})$.

\section{Hadronic Production of Radion-Higgs pair}
\label{sec:production}

The leading-order sub-processes involved in the hadronic collisions
for $p p (\bar p) \to h\phi$ are
\beq
\label{eq:subprocess}
 q \bar q \; \to \; \grav\;   \to\;  h \phi
\,, \quad
 g g \; \to \;  \grav  \; \to \;  h \phi \;.
\eeq
It is clear from Eq.~(\ref{eq:sig-gam})
that the gluon fusion process through KK graviton resonances
is dominant for
the associated production of the radion
with the Higgs boson at hadronic colliders,
especially at the LHC.
Other subprocesses such as $q\bar q \to h^*(\phi^*) \to h \phi$
are suppressed by the small Yukawa couplings.
The other processes of
$gg \to h^*(\phi^*) \to h \phi$ are also
suppressed since they occur at loop level and
through virtual intermediate scalars.
This is contrary to
the subprocesses in Eq.~(\ref{eq:subprocess})
through
the graviton poles where the majority of the cross section comes from.

The partonic cross sections for these two channels are given by
\bea
\label{eq:diff-sig-gg}
\frac{d \hat \sigma}{d \cos\theta^*} (g g \to \grav \to h \phi) &=&
\frac{\hat{g}^2_{Gh\phi}}{512 \,\pi }
\frac{\lambda^{5/2}}{\hat{s}}
\left(\frac{\hat{s}}{  \hat \Lambda_W^2}\right)^2 \,
| {\mathcal D}_G|^2
 \sin^4 \theta^* ~, \nonumber \\
 \no
\frac{d \hat \sigma}{d \cos\theta^*} (q\bar q \to \grav \to h \phi) &=&
\frac{\hat{g}^2_{Gh\phi}  }{768 \,\pi}
\frac{\lambda^{5/2}}{\hat{s}}
 \left(
    \frac{\hat s}{\hat \Lambda_W^2} \right )^2\,
 \left|  {\mathcal D}_G \right |^2 \,
 \sin^2 \theta^* \cos^2 \theta^* ~,
\eea
where $\theta^*$ is the scattering angle
in the incoming parton c.m. frame,
$\lambda = ( 1-m^2_h/\hat s - m^2_\phi/\hat s )^2
- 4 (m_h^2/\hat s) (m^2_\phi/ \hat s)$
and ${\mathcal D}_G$ is the KK graviton propagation factor,
defined by Eq.~(\ref{eq:DG}).
In the above equations, we use the Breit-Wigner prescription for the
graviton propagator.
When the c.m. energy is
away from the graviton pole, the effect of the graviton width is negligible.

\begin{figure}[t!]
\centering
\includegraphics{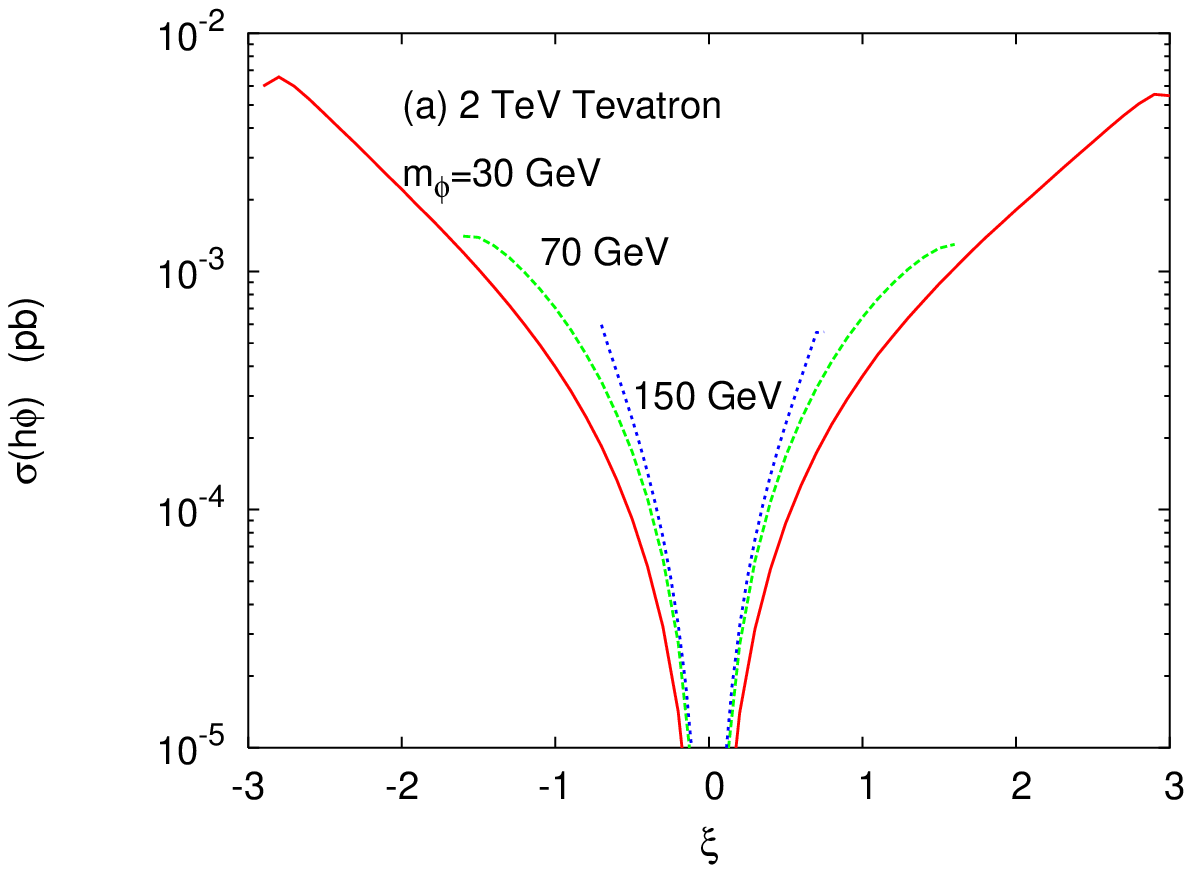}
\includegraphics{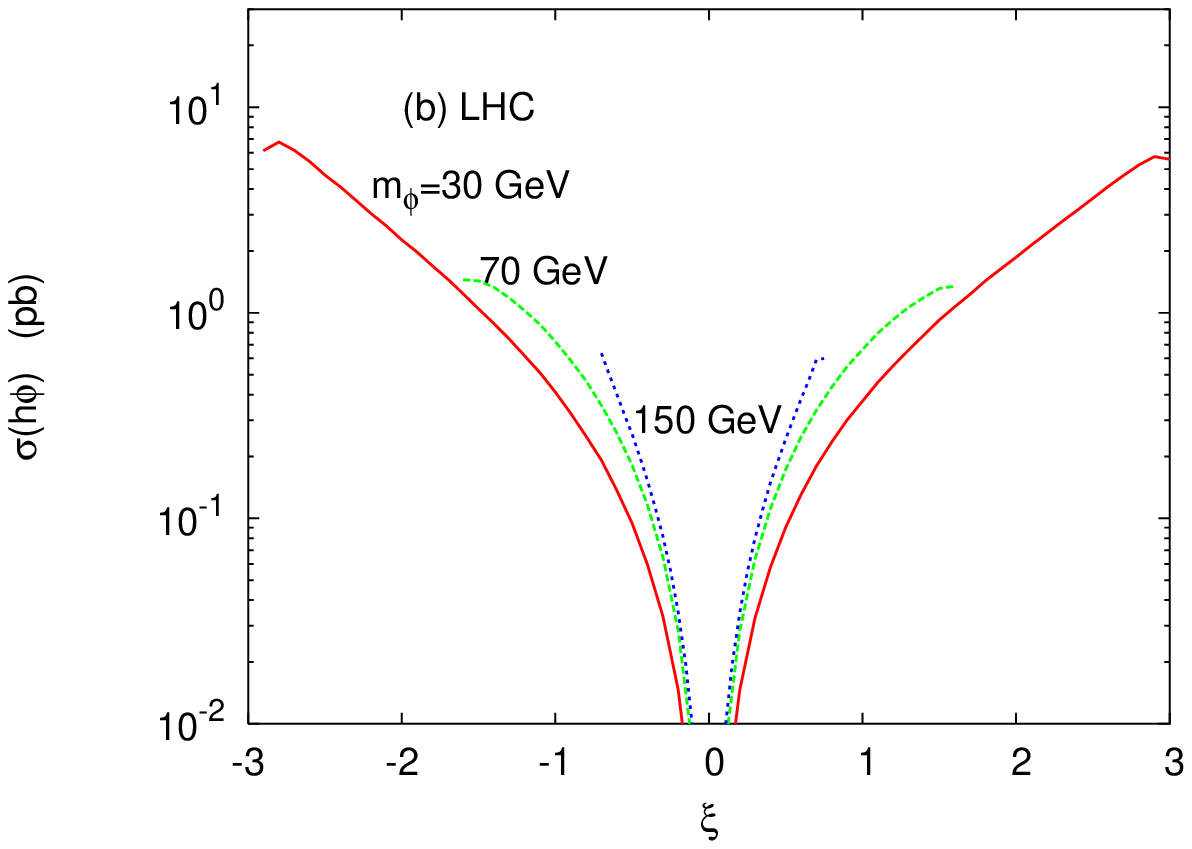}
\caption{\label{fig:tot} For $m_\phi=30,~70,~150$ GeV,
total cross section for the associated production of the radion with the
Higgs boson at (a) the 2 TeV Tevatron ($p\bar p$ collision) and (b)
at the LHC ($pp$ collision at $\sqrt{s}=14$ TeV).
As mentioned earlier, we always set $m_h=120$ GeV.}
\end{figure}

In Fig.~\ref{fig:tot},
we plot the total cross section at the Tevatron ($\sqrt{s}=2$ TeV)
and at the LHC ($\sqrt{s}=14$ TeV).
%
%
In general, the cross section at the Tevatron is of order of fb, which
means that we need a high luminosity option of the Tevatron in order to
see the radion-Higgs mixing precisely.  Fortunately, the background at the Tevatron
can be reduced substantially without hurting the signal much (we shall
show it in the next section).
At a first glance, the situation at the LHC would be
better:
As can be seen in Fig.~\ref{fig:tot}(b) the signal cross
section increases by three orders of magnitude.
For the lighter radion ($e.g.$, $m_\phi=30$ GeV) case,
the cross section can reach above the pb level.
Even for heavier radions with $m_\phi=150$ GeV,
this rare process can produce a cross section close to the pb level
if $\xi$ is sizable.
However, one has
to bear in mind that the QCD background increases more rapidly, as well as
one has to take into account other backgrounds that were small at the
Tevatron but large at the LHC.
Therefore, we anticipate that Tevatron is in fact a better place than
the LHC to search for the radion-Higgs mixing via the
$h\phi \to b\bar bjj$ final state.
In the next section, we show a detailed signal-background analysis of searching
for the $h\phi$ mixing at the Tevatron using the $h\phi \to b\bar bjj$ decay mode.

\begin{figure}[t!]
\includegraphics{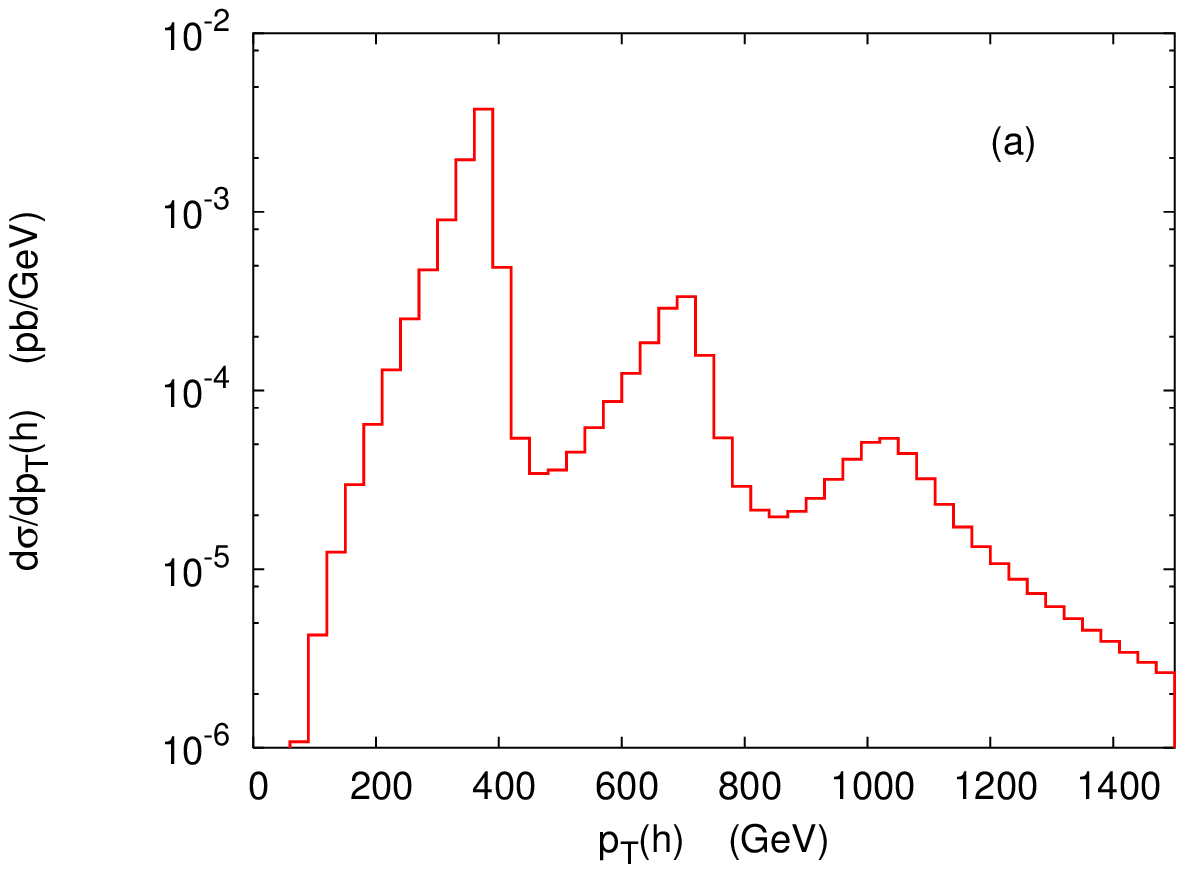}
\includegraphics{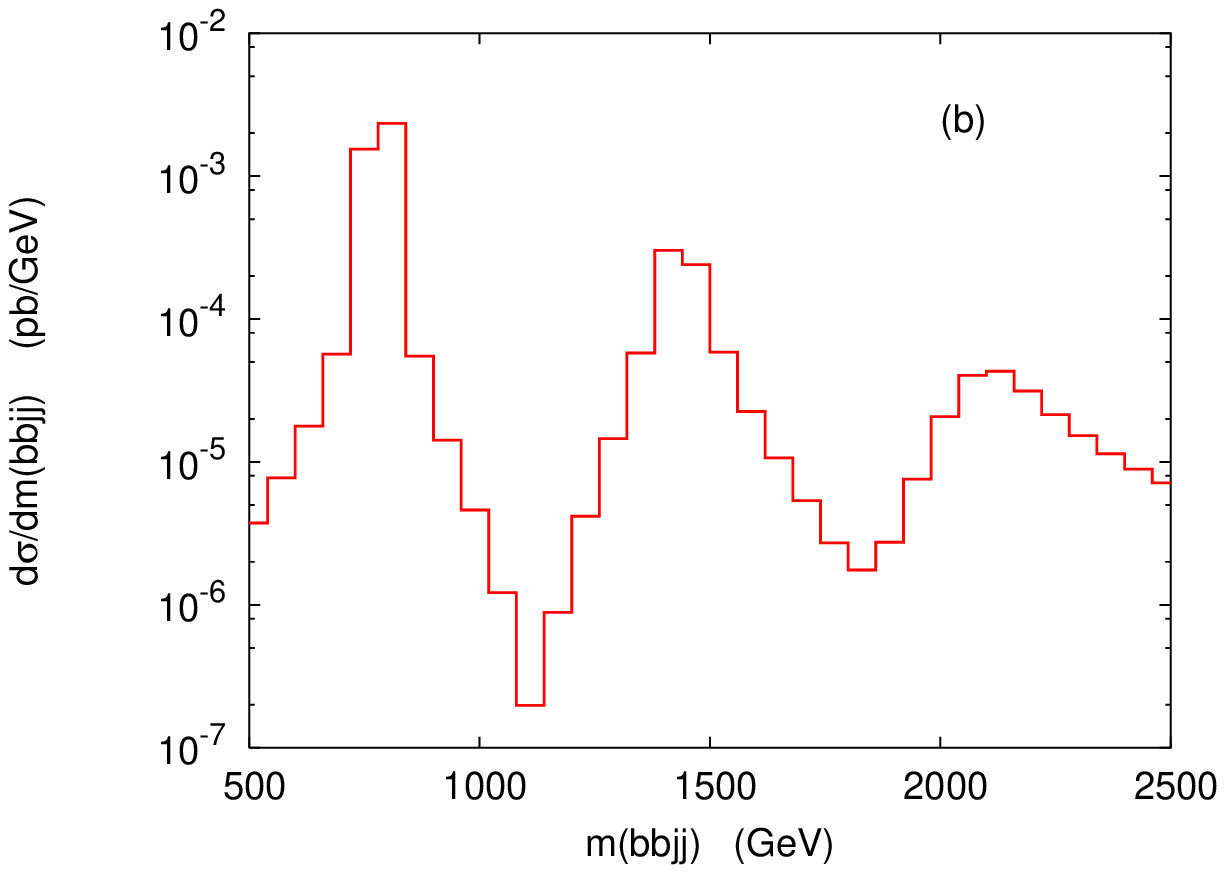}
\caption{\label{fig:ptIM} (a) Transverse momentum spectrum of the Higgs boson
reconstructed from $h\to b \bar b$, and (b) the invariant mass
$m(b\bar b jj)$ spectrum of the $h\phi$ system.
We set
$\lphi = 5~{\rm TeV}$,
${m_0}/{M_{\rm pl}} =0.1$,
$\xi=1$
$m_h = 120~{\rm GeV}$, and
$m_\phi = 30~{\rm GeV}$.
We have applied a smearing $\Delta E/E =0.5/\sqrt{E}$, where $E$ in GeV,
to all the final state particles, and
we have imposed cuts $p_T(b,j) > 20$ GeV.}
\end{figure}

For instructional purpose, we show the transverse momentum and invariant
mass distributions in Fig.~\ref{fig:ptIM}.
The resonance structure due to KK graviton states is clear in both
$p_T$ and invariant mass distributions.

\section{Decays and Detection of the Radion-Higgs pair}
\label{sec:detection}

In this section, we consider the feasibility of detecting $h\phi$
pair production in the Run II at the Tevatron.
For a Higgs boson of mass around $120$ GeV,
the major decay mode is into $b\bar b$.
The partial decay rate into $WW$
will begin to grow at $m_H \agt 140$ GeV.
Therefore, we shall focus on the
$b\bar b$ mode for the Higgs boson decay.
For a light radion with $m_\phi \lsim 2 m_W$,
because of the QCD trace anomaly,
the major decay mode of the radion is
$gg$, followed by $b\bar b$ (a distant second).
When the radion mass
gets above the $WW$ threshold, the $WW$ mode becomes dominant.
At the
Tevatron, we only consider the light radion because the production cross
section of the heavy radion is very small.
Considering the following model parameters
\beq
\lphi = 5~{\rm TeV},\quad
\frac{m_0}{M_{\rm pl}} =0.1,\quad
m_h = 120~{\rm GeV},\quad
m_\phi = 70~{\rm GeV}, \quad \xi = 1.5
\,,
\eeq
the production of the $h\phi$ pair will dominantly decay into $b\bar b g g$.
The search for this final state is similar to the search for the
associated production of $WH$ and $ZH$ with the hadronic decay of
$W$ and $Z$ performed at CDF\,\cite{cdf-wh}.  We can follow their strategies
to reduce the background.

\begin{figure}[t!]
\includegraphics[scale=0.7]{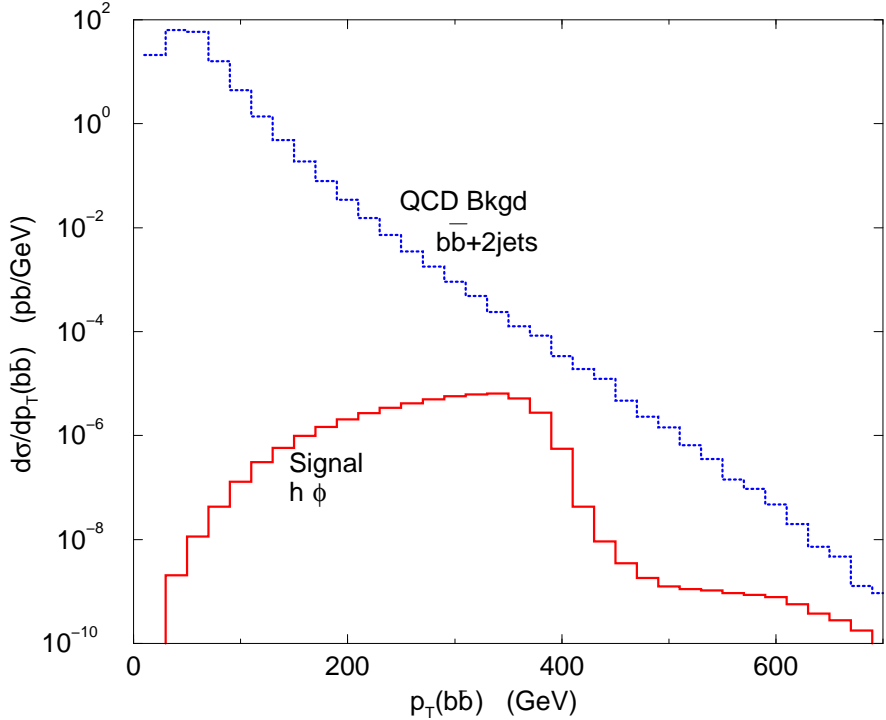}
\caption{\label{fig:ptbb} The transverse momentum
distribution of the $b\bar b$ pair of the
signal and the QCD background at Tevatron with $\sqrt{s}=2$ TeV.
We set
$\lphi = 5~{\rm TeV}$,
${m_0}/{M_{\rm pl}} =0.1$,
$m_h = 120~{\rm GeV}$, and
$m_\phi = 70~{\rm GeV}$, and $\xi=1.5$.
The imposed cuts are
$p_T(b,j) > 15$ GeV, $|y(b,j)|< 2.5$, and $\Delta R(k,l) >0.4$, where
$k,l=b,j$. }
\end{figure}

The major background comes from the QCD heavy-flavor production of
$b\bar b/c\bar c$ plus jets.
Here the $c\bar c$ pair can also fake the $b$-tagging with a lower
probability than the $b$ quark.
We calculate the QCD $b\bar b + 2$ jet background by a
parton-level calculation, in which the sub-processes are generated by
MADGRAPH\,\cite{madgraph}.
Typical cuts on detecting
the $b$-jets and light jets are applied:
\begin{eqnarray}
p_T(b) > 15 \;{\rm GeV}, &&\; p_T(j) > 15 \; {\rm GeV}, \nonumber \\
|y(b)|< 2.5,&&\; |y(j)|<2.5,\; \nonumber \\
\Delta R(k,l) > 0.4 \;\; && \mbox{where $k,l=b,j$} \;.\nonumber
\end{eqnarray}
We have applied a Gaussian smearing
$\Delta E/E = 0.5/\sqrt{E/{\rm GeV}}$ to the final-state
$b$-jets and light jets, in order to simulate the detector resolution.
Since the Higgs boson is produced together with
a radion mainly via an intermediate graviton KK state,
the Higgs boson tends
to have a large $p_T \sim m_{G^{(1)}}/2$.
Therefore, a
transverse momentum cut on the $b\bar b$ pair is
very efficient against the QCD background while only hurts the signal
marginally.
Figure \ref{fig:ptbb} shows the $p_T(b\bar b)$ distribution.
We shall apply a cut
\begin{equation}
\label{eq:ptcut}
p_T(b\bar b) > 250 \;{\rm GeV}
\end{equation}
to reduce the background.  The $b\bar b +2$ jet
background is reduced to the $0.1\%$ level.
There are other backgrounds such as  $t\bar t$
production, $Z+$jets with $Z \to b\bar b, c\bar c$,
$W b\bar b, Wc\bar c, Zb\bar b, Zc\bar c$, and
diboson and single top production, which only make up to about 1\% of
the total background.

The background is still $2-3$ orders of magnitude larger than the signal.
We further impose the mass cut on the $b\bar b$ and $jj$ pair by requiring
their invariant masses being close to the Higgs and radion masses,
respectively:
\begin{equation}
\label{masscut}
| m(b\bar b) - m_H|< 10 \;{\rm GeV},\;\; |m(jj) - m_\phi| < 10\; {\rm GeV} \;.
\end{equation}
The background is reduced to the same level as the signal:
The signal
cross section is about $0.66$ fb
while the background cross section is about
$0.94$ fb (which may vary $\sim 50\%$ due to changes in the renormalization
scale).  Further stringent cuts may help improve the signal-to-background
ratio, but it would suppress the signal event rate to an unobservable level
even at a $20$ fb$^{-1}$ luminosity.

\begin{figure}[t!]
\includegraphics[scale=0.7]{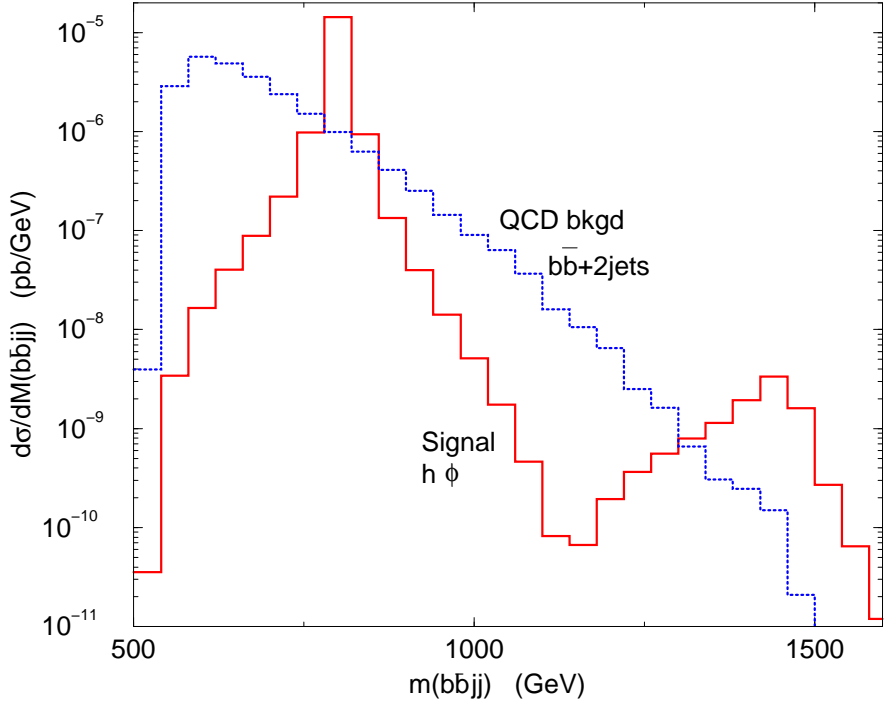}
\caption{\label{fig:inv-mass}
The invariant mass distribution of the $b\bar b jj$ of the signal
and the QCD background after applying all the cuts described in
the previous figure caption, Eq.~(\ref{eq:ptcut}), and Eq.~(\ref{masscut}). }
\end{figure}

The mass cuts of Eq.~(\ref{masscut}) can be imposed because the Higgs boson
and the radion should have been observed before we search for their mixing
effects.  We can now look at the invariant mass distribution of the
Higgs boson and the radion.  We show it in Fig. \ref{fig:inv-mass}.  It is
easy to see the peaks due to the graviton KK states.

So far we do not consider the signal-background analysis at the LHC.  The
obvious reason is that the situation at the LHC is actually getting
worse.  Although the signal is increasing by three orders of magnitude,
the background grows much faster.  Not only the $b\bar b jj$ background
that we considered, we also have to consider other
backgrounds such as $t\bar t$, $Wb\bar b$, and $Zb\bar b$,
because they are no longer negligible at the LHC.
Therefore, we anticipate that Tevatron is in fact a better place than
the LHC to search for the mixing via the $h\phi \to b\bar bjj$ final state.
If we could not find any evidence for radion-Higgs mixing in the mass
range (low to intermediate Higgs and radion mass) that we are considering
at the Tevatron, it would be even harder
to do so at the LHC.  Unless, if one looks into another mass
range of the Higgs boson and radion, say, when the radion is heavier than $2
m_Z$, then the golden mode $\phi \to ZZ \to 4l$ becomes very accessible.
In this case, the LHC would be a good place to search for the mixing.

\section{Conclusions}
\label{sec:conclusion}

In the original Randall-Sundrum scenario where all the SM
fields are confined on the visible brane,
the radion-Higgs mixing is weakly suppressed
by the radion VEV at the electroweak scale.
We have studied the phenomenological signatures of this radion-Higgs mixing
at hadron colliders.
High energy processes exclusively allowed for non-zero mixing
have been shown to provide complementary and valuable information for the mixing.
In particular,
the vertex of $\grav$-$h$-$\phi$,
one of four triple-vertices which
would vanish without the radion-Higgs mixing,
is expected to have the largest strength
in the limit of the large radion VEV,
$i.e.$, $v_0 \ll \lphi$, as suggested by the electroweak precision data.

We have studied all the partial decay widths of the KK gravitons in
the presence of the radion-Higgs mixing.
The decay mode into a gluon pair is dominant.
If the radion-Higgs mixing parameter $\xi$ is of order one,
the decay rate into a radion and a Higgs boson
becomes as large as that into a Higgs boson pair,
with the branching ratio of order $10^{-3}$.

At hadron colliders, it is feasible to produce the KK graviton resonances,
followed by their decay into a radion and a Higgs boson,
which is a clean signatures for the radion-Higgs mixing.
We have performed a signal-background analysis at the 2 TeV Tevatron,
restricting ourselves to the intermediate mass range for the Higgs boson and
the radion (otherwise the cross section at the Tevatron would be too
small to start with).
The dominant decay mode is the $h \phi \to b\bar b gg$.
Using the strategic cuts that we devised in this work, we
have been able to reduce the major QCD background
of $b\bar{b} +2j$ production
to the same level as the signal.
With an integrated luminosity of 20 fb$^{-1}$ that one can hope for the
Run II, one may be able to see a handful of such events.
We anticipate the situation at the LHC is not improving for this
intermediate mass range of Higgs and radion, because the QCD
background  increases much
faster than the signal.
On the other hand, if one looks into the heavier mass
range of the radion, say, when the radion is heavier than
$2 m_Z$, then the golden mode of
$\phi \to ZZ \to 4l$ becomes very accessible.
In this case, the final state would consist of
$h\phi \to b \bar b \ell^+ \ell^- \ell^+ \ell^-$, which is very striking.
Then the LHC would be a good place to search for the mixing.

\begin{acknowledgments}
K.C. was supported in part by the NSC, Taiwan.
The work of C.S.K. was supported by Grant No. R02-2003-000-10050-0 from BRP of the KOSEF.
The work of JS is supported by the faculty research fund of Konkuk
University in 2003.
\end{acknowledgments}

\appendix
\section{Feynman Rules}

\begin{figure}
\begin{center}
\begin{picture}(700,50)(10,0)
\Text(15,45)[]{$\phi$} \Text(15,25)[]{$q$} \Text(65,65)[]{$h$}
\Text(65,45)[]{$k_1$} \Text(65,25)[]{$\phi$} \Text(65,5)[]{$k_2$}
\DashArrowLine(15,35)(40,35){3} \DashArrowLine(40,35)(65,55){3}
\DashArrowLine(40,35)(65,15){3}
\Text(80,35)[l]{$i\frac{q^2}{\lphi}\hat{g}_\phi\,,$}
\Text(165,45)[]{$h$} \Text(165,25)[]{$q$} \Text(215,65)[]{$h$}
\Text(215,45)[]{$k_1$} \Text(215,25)[]{$\phi$}
\Text(215,5)[]{$k_2$} \DashArrowLine(190,35)(165,35){3}
\DashArrowLine(190,35)(215,55){3}
\DashArrowLine(190,35)(215,15){3}
\Text(230,35)[l]{$i\frac{q^2}{\lphi}\hat{g}_h\,,$}
\end{picture}\\
\end{center}
\smallskip
\begin{center}
\begin{picture}(700,50)(10,0)
\Text(15,45)[]{$h^n_{\mu\nu}$} \Text(15,25)[]{$q$}
\Text(65,65)[]{$h$} \Text(65,45)[]{$k_1$} \Text(65,25)[]{$\phi$}
\Text(65,5)[]{$k_2$} \DashArrowLine(40,35)(15,35){3}
\DashArrowLine(40,35)(65,55){3} \Photon(40,35)(15,35){2}{3}
\DashArrowLine(40,35)(65,15){3} \Text(80,35)[l]{$i
\,\hat{g}_{_{G\phi h}} \frac{2 k_{1\mu}k_{2\nu}}{\lwh}\,,$}
\Text(175,45)[]{$h^n_{\mu\nu}$} \Text(175,25)[]{$q$}
\Text(215,65)[]{$\phi$} \Text(215,45)[]{$k_1$}
\Text(215,25)[]{$\phi$} \Text(215,5)[]{$k_2$}
\DashArrowLine(190,35)(165,35){3} \Photon(190,35)(165,35){2}{3}
\DashArrowLine(190,35)(215,55){3}
\DashArrowLine(190,35)(215,15){3} \Text(230,35)[l]{$i
\,\hat{g}_{_{G\phi \phi}} \frac{2 k_{1\mu}k_{2\nu}}{\lwh} \,,$}
\Text(325,45)[]{$h^n_{\mu\nu}$} \Text(325,25)[]{$q$}
\Text(365,65)[]{$h$} \Text(365,45)[]{$k_1$} \Text(365,25)[]{$h$}
\Text(365,5)[]{$k_2$} \DashArrowLine(340,35)(315,35){3}
\Photon(340,35)(315,35){2}{3} \DashArrowLine(340,35)(365,55){3}
\DashArrowLine(340,35)(365,15){3} \Text(380,35)[l]{$i
\,\hat{g}_{_{G h h}} \frac{2 k_{1\mu}k_{2\nu}}{\lwh}\,. $}
\end{picture}\\
\end{center}
\caption{ Feynman rules for the tri-linear vertices in the
scalar sector. In the $h_{\mu\nu}^n$ vertices, we have made use of
the symmetry of $h_{\mu\nu}^n$ under $\mu\leftrightarrow\nu$.}
\label{trilinearfig}
\end{figure}
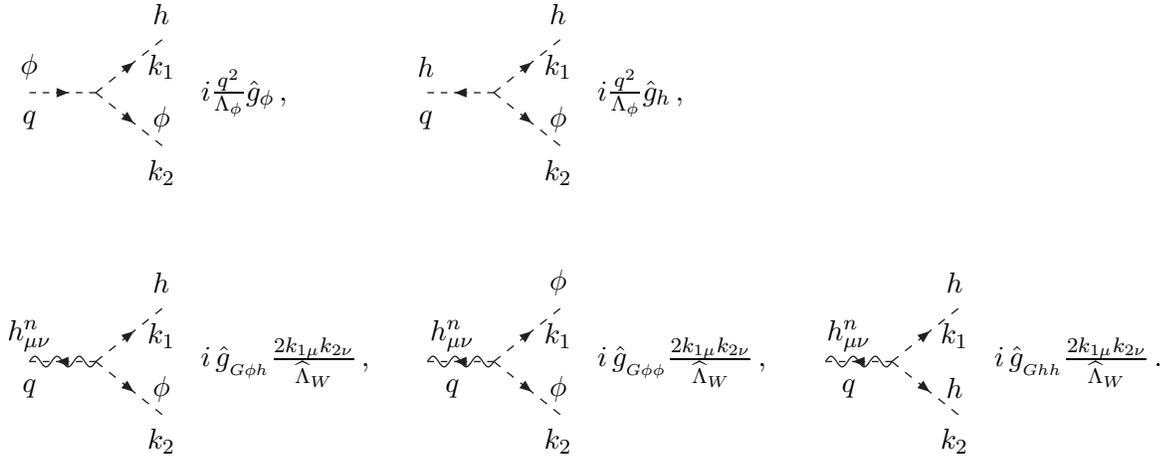

Feynman rules relevant for the
Higgs-radion production at hadron colliders
are to be summarized, focused on trilinear vertices
as depicted in
Fig.~\ref{trilinearfig}.
Properly normalized vertex factors are
\bea
\label{eq:ghat}
\hat{g}_{G h\phi} &=&
6\gam\xi\left[a(\gam b+d)+bc\right]+cd \,,
\\ \no
\hat{g}_{G \phi \phi} &=&
6 a \gam\xi\left[a\gam+2c\right]+ c^2 \,,
\\ \no
\hat{g}_{G h h} &=&
6 b\gam\xi \left[b \gam +2d\right]+d^2
\,.
\eea
In the limit of $\xi \to 0$, we have
\beq
\lim_{\xi\to 0} \hat{g}_{_{G h h}} =1,
\quad
\lim_{\xi\to 0}
\hat{g}_{_{G \phi h}} =0,
\quad
\lim_{\xi\to 0}
\hat{g}_{_{G \phi
\phi}} =0 .
\eeq
The $h-h-\phi$ and $\phi-h-\phi$ vertices involve
$q^2$-dependent couplings $\hat{g}_{h,\phi}$, parameterized by
\bea \hat{g}_{\phi} &=& \hat{g}_{\phi 1}(1+\mu_\phi^2) +
\hat{g}_{\phi2} \,\mu_h^2 -\hat{g}_{\phi 3} \,\mu_{h_0}^2 \,,
\\ \no
\hat{g}_{h} &=& \hat{g}_{h1}(1+\mu_h^2)+ \hat{g}_{h2}
\,\mu_\phi^2 -\hat{g}_{h3} \,\mu_{h_0}^2 \,, \eea where
$\mu_{h,\phi} \equiv m_{h,\phi}/\sqrt{q^2}$ ($\mu_{h_0,\phi_0}
\equiv m_{h_0,\phi_0}/\sqrt{q^2}$), and \bea \hat{g}_{h1} &=& 6 b
\xi \Bigl\{\gam ( a d + bc) +c d\Bigr\} + a d^2 \,, \quad\quad\;
\hat{g}_{\phi 1} = 6 a \xi \Bigl\{\gam ( a d + bc) +c d\Bigr\} + b
c^2 \,,
\\ \no
\hat{g}_{h2} &=& d \left\{ 12 a b \gam \xi +2 b c + a d ( 6
\xi-1)\right\} \,,\quad \hat{g}_{\phi 2} =c \left\{12 a b \gam \xi
+2 a d  + b c ( 6 \xi-1)\right\} \,,
\\ \no
\hat{g}_{h3} &=& 4 d ( a d + 2 b c ) + 3 \gam^{-1} c d^2
\,,\quad\quad\quad\quad~ \hat{g}_{\phi 3} = 4 c (2 a d + b c )+ 3
\gam^{-1} c^2 d \,. \eea

\begin{figure}
\vspace*{.5in}
\begin{center}
\begin{picture}(330,50)(40,0)
\Text(15,45)[]{$h^\n_{\mu\nu}$} \Text(15,25)[]{$q$}
\Text(65,65)[]{$A_\rho^a(q_1)$}
\Text(65,5)[]{$A_\sigma^b(q_2)$}
\Vertex(40,35){3} \DashLine(40,35)(15,35){3}
\Gluon(40,35)(65,55){2}{4} \Gluon(40,35)(65,15){2}{4}
\Text(85,35)[l]{$-\frac{i}{\lwh}\delta^{qb} \left[ q_1\cdot q_2
C_{\mu\nu,\rho,\sigma}+ D_{\mu\nu,\rho\sigma}(q_1,q_2) \right]$}
\end{picture}\\
\end{center}
\vspace*{.5in}
\begin{center}
\begin{picture}(330,50)(40,0)
\Text(15,45)[]{$\phi,h$} \Text(15,25)[]{$q$}
\Text(65,65)[]{$A_\rho^a(q_1)$}
\Text(65,5)[]{$A_\sigma^b(q_2)$}
\Vertex(40,35){3} \DashLine(40,35)(15,35){3}
\Gluon(40,35)(65,55){2}{4} \Gluon(40,35)(65,15){2}{4}
\Text(85,35)[l]{$i\frac{\hat{c}_{\phi,h}}{v} \,\delta^{ab} \left[
q_1\cdot q_2 \,\eta_{\rho\sigma}-q_{1\rho} q_{2\sigma} \right]$}
\end{picture}\\
\end{center}
\caption{ Feynman rules involving gluon pair. }
\label{vvfffig}
\end{figure}
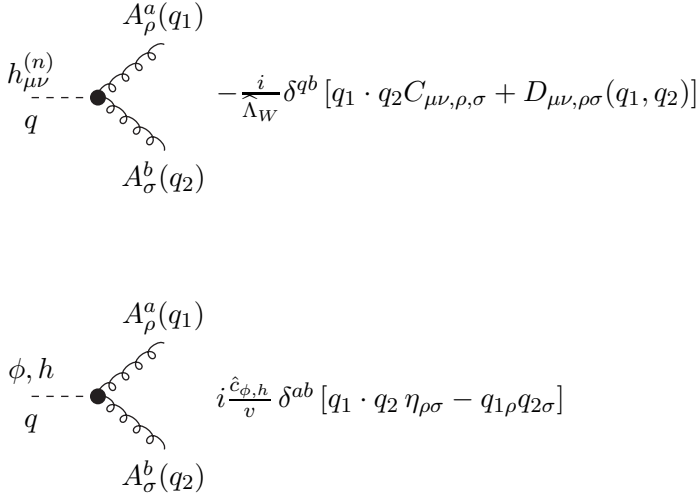

For completion, we review the Feynman rules involving a gluon pair
in Fig.~\ref{vvfffig}. We refer for the expressions of
$C_{\mu\nu,\rho\sigma}$ and $D_{\mu\nu,\rho\sigma}$ to
Ref.~\cite{Han-Zhang}, and
\bea \hat{c}_h &=& - \frac{\al_s}{4\pi}
\left[ (d+\gam b) \sum_i F_{1/2}(4 \mu^2_{ih}) - 2 b_3 \gamma b
\right] \,,
\\ \no
\hat{c}_\phi &=& - \frac{\al_s}{4\pi}
\left[ (c+\gam a) \sum_i
F_{1/2}(4 \mu^2_{i\phi}) - 2 b_3 \gamma a
\right] \,.
\eea
Here
$F_{1/2}(x)= -2x[1+(1-x)f(x)]$ and $f(x)= -(1/4)\ln \left[
(\sqrt{1-x}+1)/(\sqrt{1-x}-1) \right]^2$ with $\mu_{i j}\equiv
m_i/m_j$.

\section{Helicity amplitudes for $gg\to h \phi$ and $q\bar{q} \to h \phi$}

For the gluon fusion process of
\beq
g(q_1,\lam_1) +  g(q_2,{\lam}_2) \rightarrow h(k_1)+\phi(k_2) \,,
\eeq
four momenta in the parton c.m. frame are defined by
\bea
q_1^\mu
\!\!&=&\!\! \frac{\sqrt{\hats}}{2} \left(1,0,0,1 \right), \quad
q_2^\mu = \frac{\sqrt{\hats}}{2} \left(1,0,0,-1 \right), \\ \no
k_1^\mu \!\!&=&\!\! \frac{\sqrt{\hats}}{2}
\left(1+\mu_h^2-\mu_\phi^2,\phantom{-}\lam^{\frac{1}{2}}\sin\theta^*,
0,\phantom{-}\lam^{\frac{1}{2}}\cos\theta^* \right),
\\ \no
k_2^\mu \!\!&=&\!\! \frac{\sqrt{\hats}}{2}
\left(1-\mu_h^2+\mu_\phi^2,-\lam^{\frac{1}{2}}\sin\theta^*,
0,-\lam^{\frac{1}{2}}\cos\theta^* \right),
\eea
where
$\mu_{h,\phi} \equiv m_{h,\phi}/\sqrt{\hats}$,
$|\vec{p}_h|=|\vec{p}_\phi| =\lam^{\frac{1}{2}} \sqrt{\hats}/2$
and $\lam = 1+ \mu_h^4+\mu_\phi^4- 2\mu_h^2 -2 \mu_h^2 -2 \mu_h^2
\mu_h^2$.
The $\lam_{1(2)}=\pm$ denotes the gluon polarization,
which specifies its polarization vectors as
\beq
\es^\mu_1(q_1,\lam_1) = \frac{1}{\sqrt{2}} \left( 0,-\lam_1,-i,0
\right), \quad \es^\mu_2(q_2,\lam_2) = \frac{1}{\sqrt{2}} \left(
0,\lam_2,-i,0 \right) . \eeq

Defining the propagator factors of the KK-graviton, Higgs and
radion by
\beq
\label{eq:DG}
{\mathcal D}_G =\sum_{n=1}^\infty
\frac{\hats}{\hats - m^2_{G^{(n)}}+ i m_{G^{(n)}}\Gam_{G^{(n)}} } \,,\quad
{\mathcal D}_{h,\phi} =\frac{\hats}{\hats-m_{h,\phi}^2+ i
m_{h,\phi}\Gam_{h,\phi}} \,,
\eeq
the helicity amplitudes
$\delta^{ab} \M_{\lam_1,\lam_2}$ with the color factor of $\delta^{ab}$ are
\bea
\M_{++} &=& -
\frac{\hats}{2 v \lphi} \left( \hat{c}_h \hat{g}_h \mathcal{D}_h +
\hat{c}_\phi \hat{g}_\phi \mathcal{D}_\phi \right),
\\ \no
\M_{+-} &=& -\frac{\hats \lam}{\lwh^2}\hat{g}_{G\phi
h}\mathcal{D}_G \sin^2\theta^* \,,
\eea
where $\M_{++}=\M_{--}$ and  $\M_{+-}=\M_{-+}$ guarantied
by {\em CP} invariance.
Note that the contribution of the scalar mediation
is separated from that of KK gravitons according to
the gluon polarization.

For the differential cross section of
\beq
\frac{d \hat{\sigma} }{d \cos\theta^* } =
\frac{\lam^{1/2}}{32\pi \hats} \overline{|\M|^2}
\,,
\eeq
the gluon-polarization averaged amplitude squared
is
\beq
\overline{|\M|^2} = \frac{1}{2}
\cdot \frac{1}{2} \cdot \frac{1}{8} \cdot \frac{1}{8} \cdot 8
\left[ |\M_{++}|^2 +|\M_{+-}|^2 +|\M_{-+}|^2 +|\M_{--}|^2 \right]
\,,
\eeq
where the factor $1/2$ is for the gluon polarization
average, the factor $1/8$ for the gluon color average, and the factor $8$
for the color-sum.


For the $q\bar{q}$ annihilation production of the Higgs and radion,
the helicity amplitudes are the same as the case of $e^+ e^- \to h
\phi$ except for the color factor~\cite{ours}.
Coordinating notations, we
have
\beq
\overline{|\M|^2}(q\bar{q}\to h\phi)
=\frac{\hat{g}_{G\phi h}^2}{24} \left( \frac{\lam \hats}{\lwh^2}
\right)^2 |\mathcal {D}_G|^2 \sin^2 \theta^*\cos^2 \theta^* \,.
\eeq

\end{document}